# A New Method For Flushing of Subsea Production Systems Prior to Decommissioning or Component Disconnection

**L. C. Sevillano,** Wood Group AS; **M. Stanko\*** and **S. Sangesland,** Norwegian University of Science and Technology

*Corresponding author; email: milan.stanko@ntnu.no

## Summary

Before decommissioning, or when parts and components of a subsea production system must be replaced, the system is typically flushed with MEG. This is done to reduce the content of hydrocarbons such that harmful discharges to sea are minimized when a component is disconnected. This paper outlines a novel subsea flushing system which uses a subsea tool to improve the performance of the flushing operation. Flushing is typically performed from the host offshore facility through the existing flowlines for chemical inhibition, from a service vessel, or a combination of both. Flushing is performed along a pre-determined flow path and ensuring a minimum flow velocity inside the components to achieve efficient sweeping. It is a costly operation in terms of vessel and personnel mobilization, required amount of MEG and non-productive time. The new method outlined in this paper uses a small-diameter, high-pressure supply line and a subsea deployed tool containing a pump which recirculates the cleaning fluid through the component or system to be retrieved. The main benefit of this method when compared against conventional practices is that it allows achieving higher fluid speeds inside the subsea equipment being flushed, while injecting smaller flow rates from the surface vessel. The high fluid speeds are achieved with the recirculation pump. The higher fluid speeds ensure efficient sweeping of hydrocarbons from complex paths. A reduced flow rate from the surface vessel also allows a small diameter high pressure supply line to be used, which allows for reduced weight and storage.

The study is a numerical simulation of the method applied to a subsea jumper geometry. The injection flow rates required to achieve an efficient flushing were determined from previous experimental work. Calculations were made to estimate the pressure and power requirements for performing the flushing operation as well as the design requirements for the supply line concerning dimensions, material properties and the storage space needed on the support vessel. The performance of the proposed novel system was compared to that of conventional flushing systems. As environmental concerns increase, the presented method has the potential to make the flushing process more efficient while reducing costs associated with support vessels and the materials needed. The novel system may also be deployed using a low-cost Inspection Maintenance and Repair (IMR) vessel. The subsea tool is connected to the subsea production system, either through dedicated connection ports or using pipe clamp connectors with pipe wall penetrators.

## Introduction

Offshore oil and gas facilities worldwide must keep the hydrocarbon content of discharges to sea below the acceptable levels set by local authorities. In the North Sea, for instance, the required monthly average concentration of hydrocarbons discharged in effluents, such as produced water, must be below 30 mg/l (Carpenter 2019). The same requirement applies to discharges from a subsea production systems (SPS), for example, when replacing and disconnecting components in connection with decommissioning, replacement, and maintenance. To achieve a low concentration of hydrocarbon in subsea components before disconnection, components are cleaned or flushed before retrieval to surface. In some cases, it is not possible to reduce the hydrocarbon content of the component by flushing, which requires applying for especial permits and payment of penalties to the authorities.

Subsea pipelines may be cleaned and flushed with seawater (Yap 2018), but when parts and components of a subsea system must be replaced, for example jumpers, manifolds, multiphase meters and pumping modules, the system is typically flushed with Mono-ethylene glycol (MEG) or Nitrogen to minimize flow assurance issues such as the formation of hydrates and corrosion. MEG is included in the "List of Substances Used and Discharged Offshore which Are Considered to Pose Little or No Risk to the Environment" (PLONOR) prepared by the OSPAR (Convention for the Protection of the Marine Environment of the North-East Atlantic). Therefore, it is possible to discharge without permission (but reporting is often required). Despite this, oil and gas companies do not release MEG unless necessary because it is costly.

Flushing is typically performed from the host offshore facility through the existing flowlines for chemical inhibition (that are usually part of the umbilical), from a ship, or a combination of both. It may require a specialized flushing tool to be developed, such as the tree running flushing tool presented by Yap (2018). A literature review of past extraordinary applications for emission permissions to the Norwegian Environment Agency and yearly environmental reports of oil and gas companies operating in Norway seem to indicate that flushing operations are usually planned with the goal of zero release of hydrocarbons to Sea. However, the flushing method will most likely leave some hydrocarbon trapped inside the geometry, for example in recesses and high points. This doesn't necessarily indicate that the remaining hydrocarbons will be released to sea when the component is retrieved, since it is usually difficult for the remaining small amounts of hydrocarbons to flow out of the geometry.

It is not uncommon that subsea equipment consists of complex pipe geometries (interior of valves, rotating and metering equipment). Flushing these subsea components is challenging. Due to the uncertainties regarding the displacement process, the operator or service companies often perform these operations in a conservative manner, i.e., circulating for long times and at high rates, which is costly in both time and resources. The goal of this paper is to present a new concept for flushing of subsea equipment which allows to achieve lower concentrations of hydrocarbons in complex subsea components at lower flow rates of flushing fluid.

In this work, the term "flushing efficiency" refers to the volume fraction of hydrocarbons in the geometry after the flushing process. A high flushing efficiency flushing is when that volume fraction is low (close to zero). A poor flushing efficiency is when that volume fraction is high (close to 1).

**Background - Previous Studies**

The new concept outlined in this paper for flushing of subsea equipment is based on results obtained from experimental and numerical studies (Opstvedt 2016, Folde 2017, Valentini 2019). The geometry used in these works consisted of a 160 mm diameter U-shaped jumper made of transparent PVC (**Fig. 1**). The jumper was pre-filled with distilled water with 3.4%w salt or oil (Exxsol D60) and flushed with either oil (in case initially filled with water) or water (in case initially filled with oil).

Opstvedt (2016) and Folde (2017) performed experimental testing and numerical modeling of the flushing process considering the flushing fluid is injected from one end (using a centrifugal pump) and a mixture is received on the other end. The content of oil and water in the jumper was measured in time by stopping the flushing process and draining the volume. The total volume had 165 l. Several flushing rates were tested.

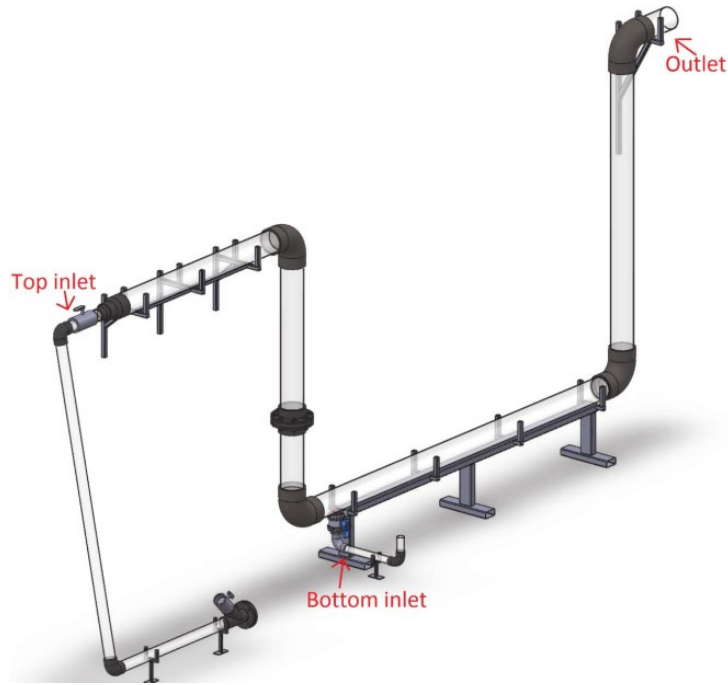

**Fig. 1—3D model of jumper geometry (Opstvedt 2016).**

**Fig. 2** presents the evolution of the oil volume fraction in the jumper with time, when initially filled with water, for 5 flushing rates of oil. The two highest rates allow to achieve a 100% concentration of oil in short time, but for the 3 lowest flushing rates, there is still significant amounts of water left in the jumper at the end of the experiments. Similar results were found for water displacing oil (**Fig. 3**).

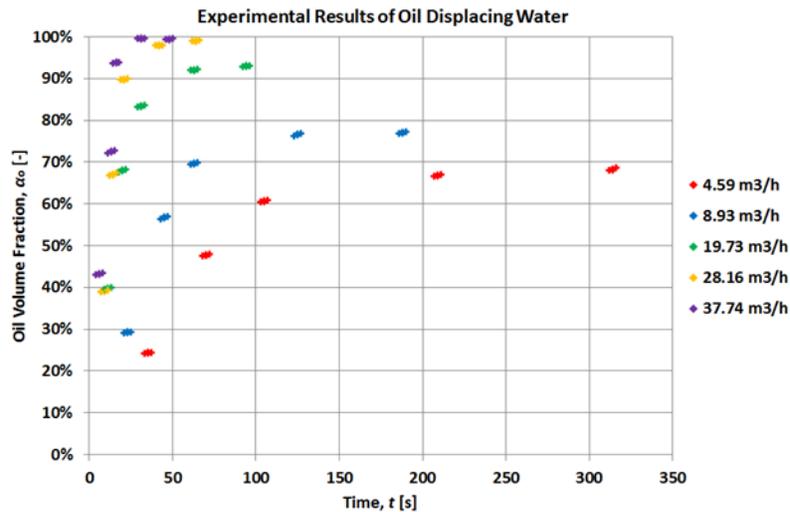

**Fig. 2—Evolution of oil volume fraction in the jumper with time, for 5 flushing rates of oil (Folde 2017).**

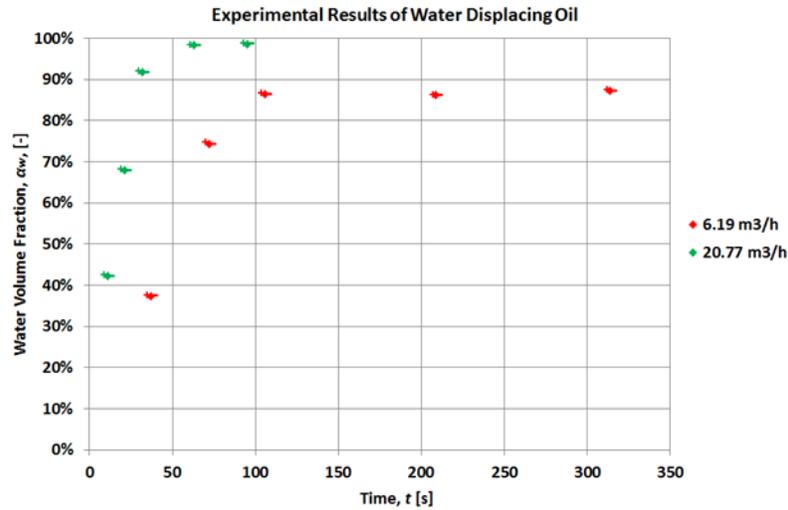

Fig. 3—Evolution of water volume fraction in the jumper with time, for 2 flushing rates of water (Folde 2017).

Opstvedt (2016) and Folde (2017) also performed numerical simulations using a commercial computational fluid dynamics software (ANSYS CFX v16.2) and a transient multiphase pipeflow simulator (Ledaflow v2.1.252.024). Results are shown in **Fig. 4** when using the transient multiphase pipeflow simulator (Folde 2017). The agreement between the simulation results and the measured values is fair.

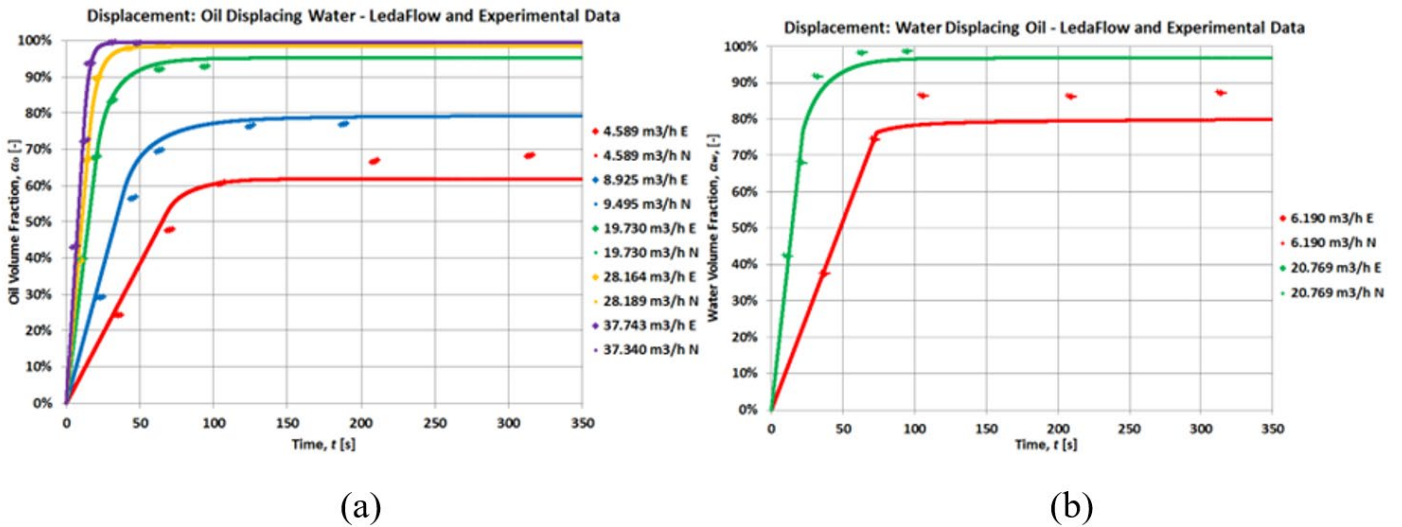

(a)                                                        (b)

Fig. 4—Comparison of numerical (solid lines) versus experimental results (points) of a) Evolution of oil volume fraction in the jumper with time, for 5 flushing rates of oil and b) Evolution of water volume fraction in the jumper with time, for 2 flushing rates of water (Folde 2017). Error bars are added to the experimental points but are too small to be visible to the naked eye.

A disadvantage of employing this flushing method (i.e. injection from one end and receiving the discharge on the other end) is that high volumetric rates of the flushing fluid are required to achieve high flushing efficiency. Additionally, after some time it won't be possible to remove any more undesirable fluid from the domain by flushing.. The leftover amount of the undesired fluid in the domain depends strongly on the flushing rate.

Valentini (2019) performed flushing experiments with a modified test loop (shown in **Fig. 5**). The inlet and outlet of the jumper were connected with a transparent pipe section of 63 mm internal diameter (ID). A flow meter and an ejector (jet pump) were installed on that additional pipe section. A simplified schematic of an ejector (jet pump) is shown in Fig. 6. The suction side of the ejector (Jet pump) was connected towards the outlet and the discharge part of the ejector (Jet pump) was connected towards the inlet. The outlet was also connected to a storage tank (separator). The ejector (Jet pump) was driven with flushing fluid (either water or oil) using a centrifugal pump. The flow rate of the bypass upstream the jet pump was monitored with a flow meter. The flow rate of the injected flushing fluid was measured with a flow meter.

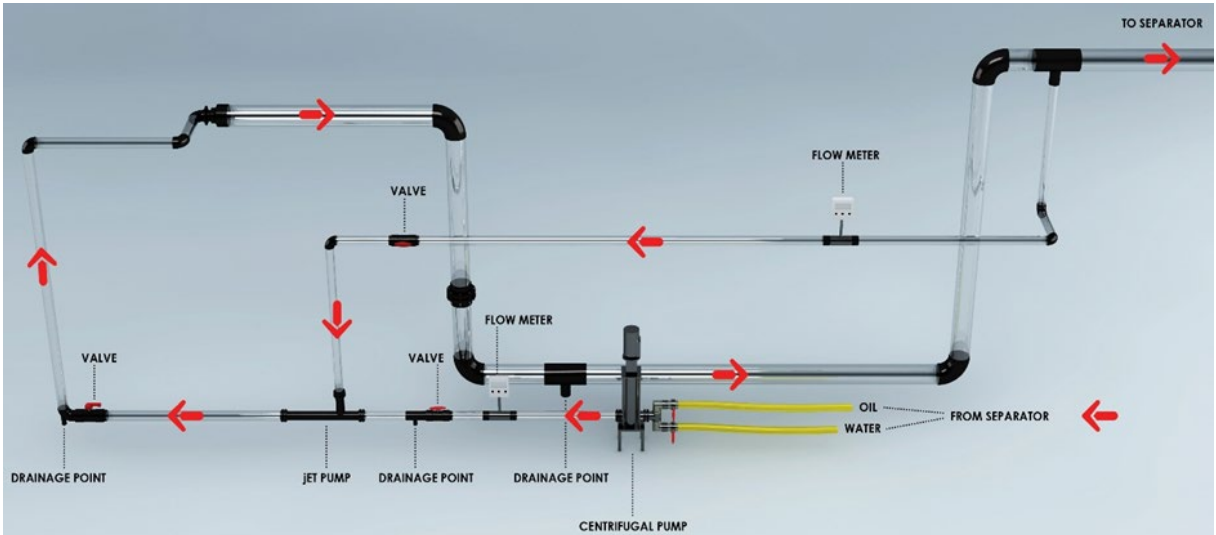

Fig. 5—3D model of jumper geometry and flushing arrangement employed by Valentini (2019).

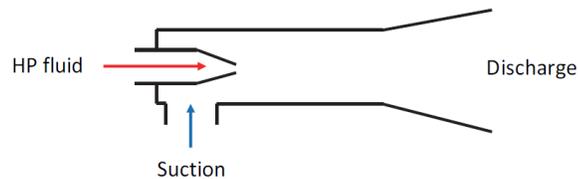

Fig. 6—Simplified schematic of a jet pump.

The oil volume fraction in the domain was estimated by stopping the flushing process, let the fluids segregate, measure the levels of oil in all pipe sections and compute total volumes of water and oil.

The working principle of this setup is that the flushing fluid is introduced into the system through the ejector (Jet pump), the ejector (Jet pump) creates some re-circulation in the jumper-bypass loop. As flushing fluid enters the system, undesired fluid plus flushing fluid leaves the system through the outlet, the geometry becomes cleaner in time.

The benefit obtained from recirculating the fluid in the system is to achieve higher flow velocities to displace the undesired fluid, but with smaller flow rates of flushing fluid being fed into the system.

The experimental results are presented in **Fig. 7** for the case of the geometry initially filled with oil and flushed with water. 3 water rates are presented (5.00, 10.00, 20.77 m³/h). The total flow rates induced in the jumper (due to the ejector (Jet pump)) are 9.80 m³/h, 20.62 m³/h and 41.40 m³/h respectively. Contrary to what was observed from the results of Folde (2017), results show that is possible to achieve a low content (below 15%) of oil in the jumper with all flushing rates of water, but that it takes longer time for low flushing rates

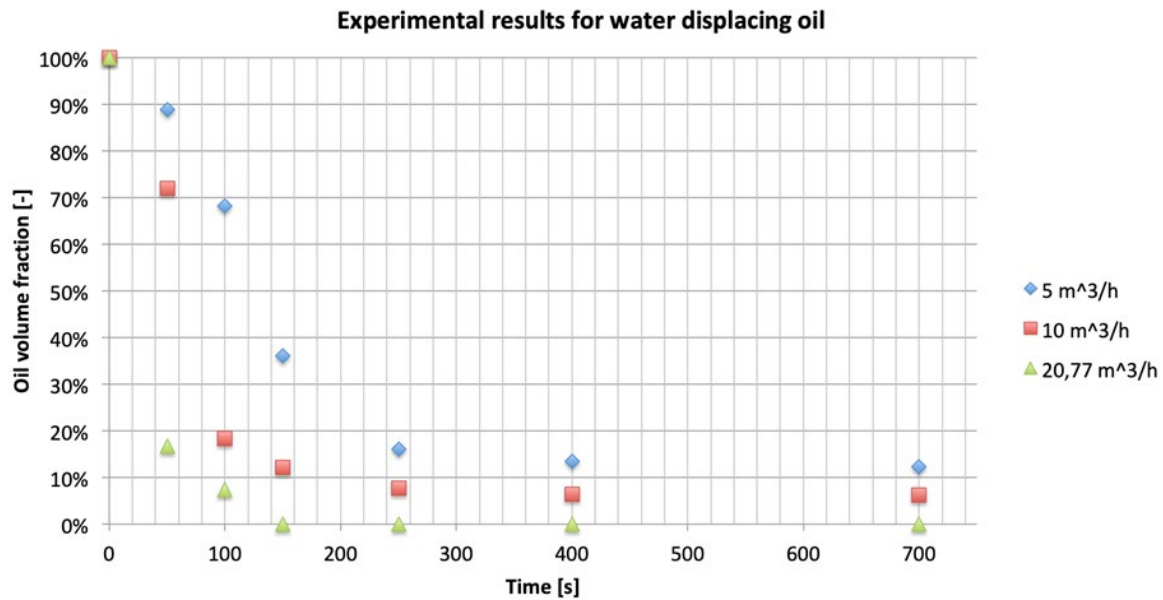

**Fig. 7—Evolution of oil volume fraction in the system with time, for 3 flushing rates of water (Valentini 2019).**

A simplified analytical model to estimate the remaining oil volume fraction in time was developed by performing a mass balance on the domain (details provided in Appendix A). The main assumption of the model is that, due to the strong recirculation, it is possible to approximate the flushing geometry as a tank with homogeneous properties (e.g. oil volume fraction). The inlet stream consists of the flushing fluid rate. The outlet stream has the same volumetric rate as the inlet, but the same oil volume fraction as the tank (assumes incompressible liquids). This model might be too simplistic for systems with large residence times such as pipelines. For those cases, a model based on axial dispersion such like the ones presented by Taylor (1954) and Austin and Palfrey (1963) could be more suitable.

The experimental setup and procedure were also simulated using the transient pipeflow simulator (LedaFlow).

**Fig. 8** shows the results of the case when flushing with 5 m$^3$/h of water (domain initially filled with oil). The experimental measurements are shown with crosses, the results obtained with the simplified mathematical model are shown in blue and the transient pipe flow simulator are shown with green. The agreement between the models and measurements is fair. It seems the simplified model provides a slightly better match overall. This model could be used for planning flushing operations, e.g. determining the flushing rates to inject and for how long to achieve a target volume fraction of hydrocarbons in the domain.

The numerical results obtained by Folde (2017) using the transient pipe flow simulator with a flushing rate of 5 m$^3$/h are also shown in Fig. 8 (solid red line). There is a significant difference between flushing using no recirculation and recirculation. The flushing with recirculation allowed to achieve a significantly lower amount of oil in the domain, although with longer recirculation time.

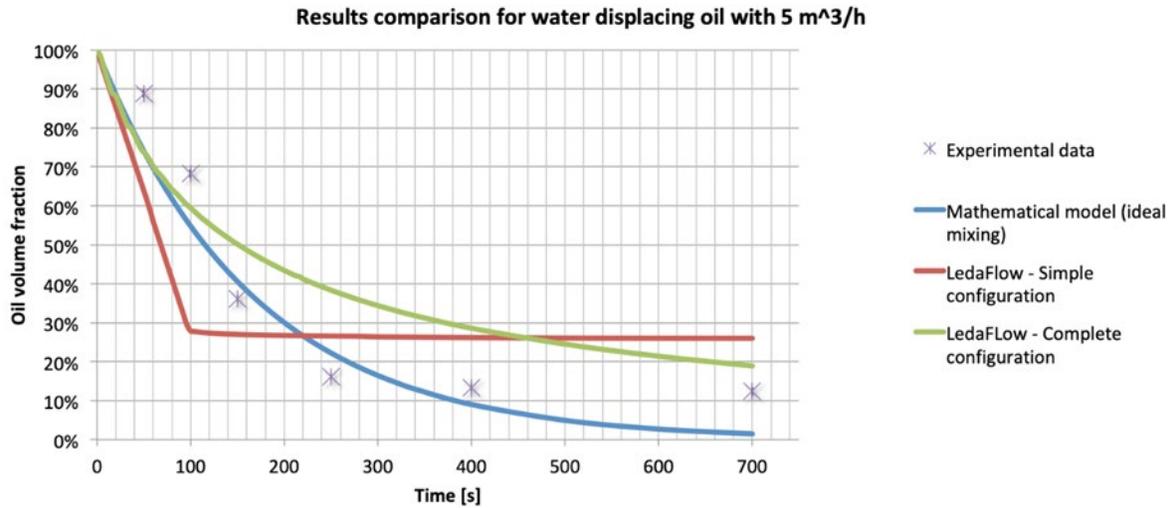

**Fig. 8—Evolution of oil volume fraction in the system with time, for a flushing rate of water equal to 5 m³/h (Valentini 2019). Model results and experimental values.**

## Possible configurations and layouts of the new flushing concept

Fig. 9 shows several possible configurations and layout of the new flushing concept. The images show a subsea module with equipment that must be retrieved, the recirculation pump, the injection port. There could be variations to these configurations proposed.

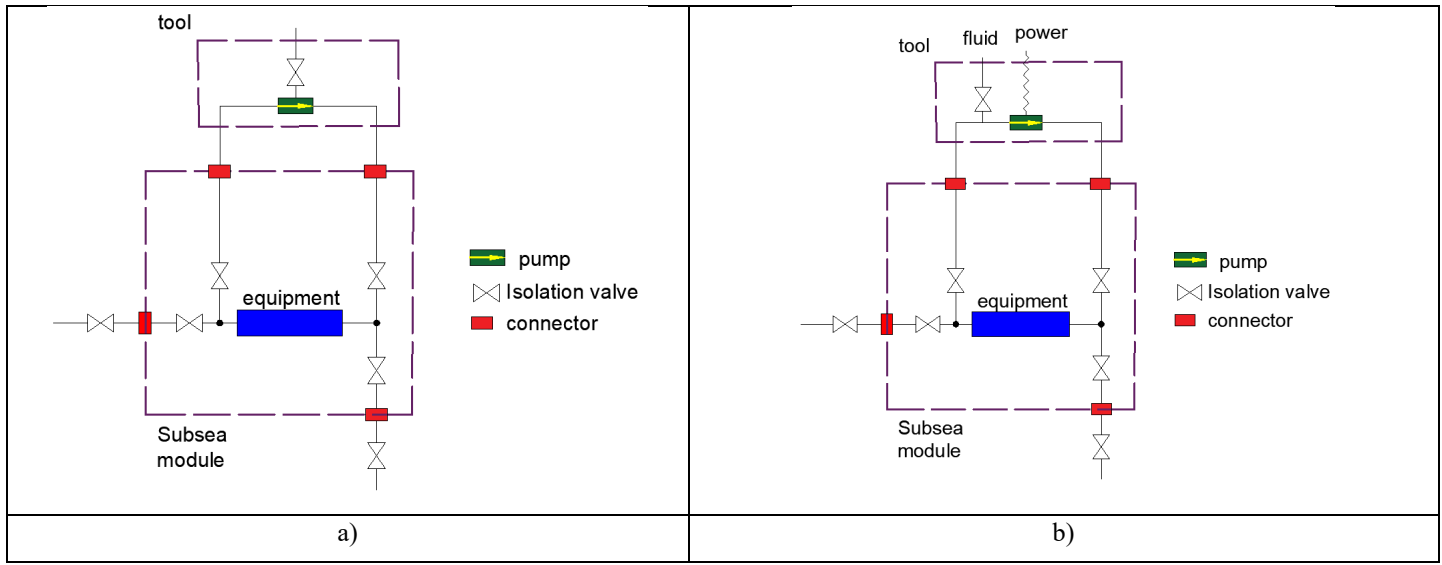

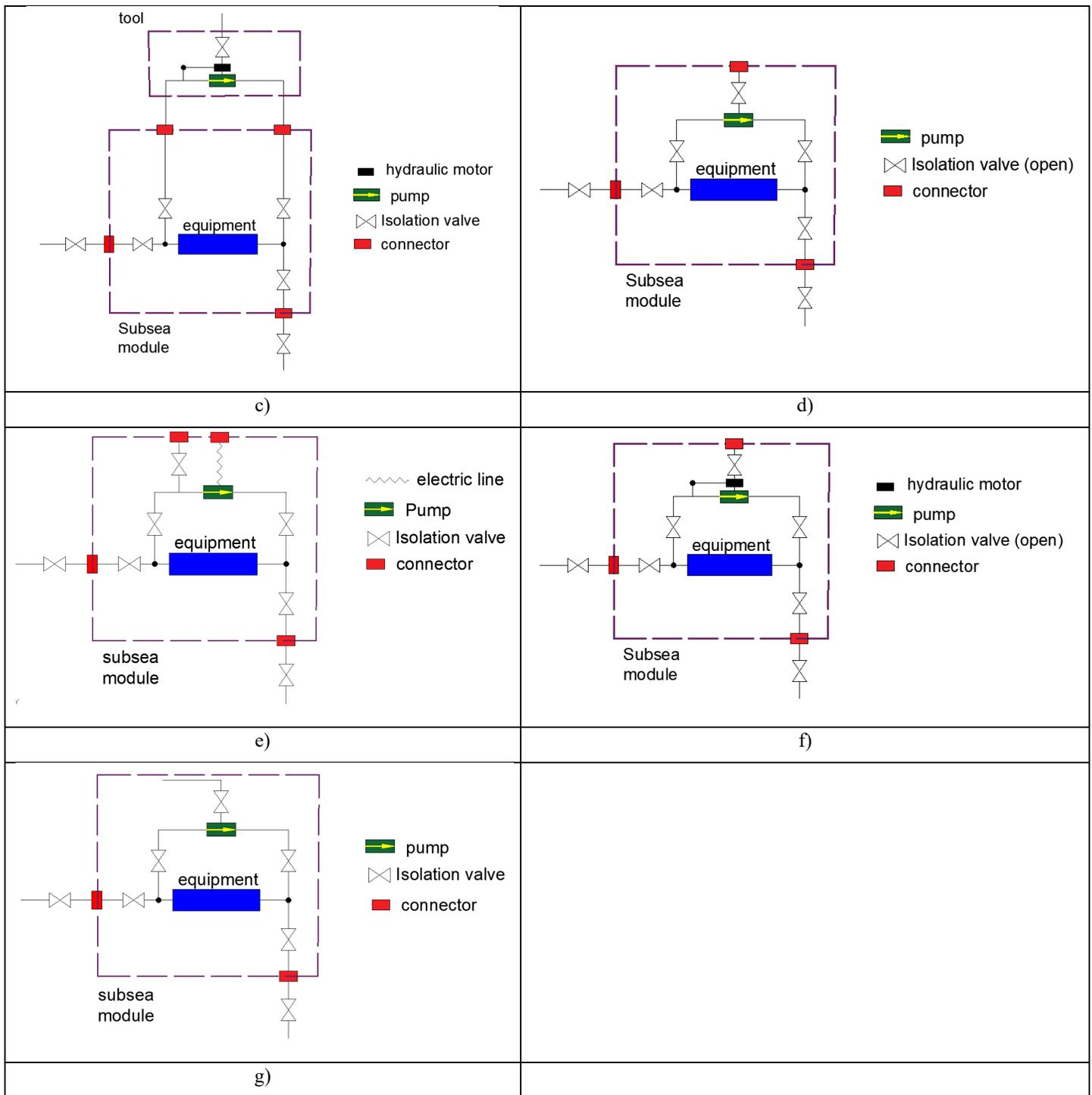

**Fig. 9 Possible configurations and layouts of the new flushing concept**

a) Flushing fluid injected from a vessel with the intervention tool. The pump is in the tool and is driven by the injection fluid (e.g. a jet pump).

b) Flushing fluid injected from a vessel with the intervention tool. The pump is in the tool and is driven with an electric motor.

c) Flushing fluid injected from a vessel with the intervention tool. The pump is in the tool and is driven with a hydraulic motor actuated by the injection fluid.

d) Flushing fluid injected from a vessel with the intervention tool. The pump is on the subsea module and driven by the injection fluid (e.g. a jet pump).

e) Flushing fluid injected from a vessel with the intervention tool. The pump is on the subsea module and driven by an electric motor powered by the intervention tool.

f) Flushing fluid injected from a vessel with the intervention tool. The pump is on the subsea module and and is driven with a hydraulic motor actuated by the injection fluid.

g) Flushing fluid injected via the chemical injection line in the subsea module. The pump is on the subsea module and driven by the injection fluid (e.g. a jet pump).

**Qualitative comparison between the proposed method and the standard flushing method**

Table 1 presents a qualitative comparison between the standard flushing method and the new method proposed for several categories. There are some items for which the traditional flushing is advantageous, while there are others where the new method is more attractive.

Table 1 – Qualitative comparison between the traditional flushing method and the new method proposed.

| Category | Traditional flushing | Recirculation-based flushing |
|---|---|---|
| **Required flushing time** | Short (minutes, hour) | Long (hours) |
| **Required injected flushing rates** | High | Low (1/2-1/4 of traditional flushing[1]) |
| **Required flushing volumes** | Big (high rates, short time) | Big (low rates, long time) |
| **Applicability for long pipelines** | Low | Low |
| **Effectiveness when dealing with complex geometries** | Medium | High |
| **When flushing from vessel, size of fluid conduit** | Big | Small |
| **When flushing from vessel, capacity of deck pump** | Big | Small |
| **When flushing from vessel, complexity of subsea tool** | Low | High |
| **When flushing from vessel, load capacity of vessel** | Big | Low |
| **Compatibility with existing subsea systems** | High | Medium (access ports to establish recirculation are required) |
| **Risk of small amounts of hydrocarbons leaking out of geometry when disconnecting** | Low | High[1] |

1. Based on the experimental data, to achieve the same value of hydrocarbon volume fraction in the domain at the end of the flushing time.

2. The new method proposed is based on homogenizing (mixing) hydrocarbons with the flushing fluid (by means of a pump) and then gradually replacing the hydrocarbons by injecting flushing fluids. So, at the end of the flushing time, the geometry will most likely be filled with flushing fluid with traces of hydrocarbon. If the disconnection of the equipment is done right afterwards the flushing is completed, without letting it segregate (MEG is heavier than hydrocarbons), there is a chance there will be some hydrocarbon leaking out of the geometry together with the MEG. The changes of this happening seem to be higher for our method than for the standard single pass flushing method. Workarounds to address this situation could be to wait sometime after the flushing is completed, or to circulate for longer time, to reduce the amount of hydrocarbon to a minimum.

**Cost comparison**

It is challenging to compare both methods with respect to cost, since there are several aspects to consider, and many are uncertain. Therefore, some general comments will be provided that could affect cost:

When flushing using a service vessel, pump located in the intervention tool

- The new method will require longer flushing time than the traditional method, which represents higher costs due to vessel rent rates and higher risk (the vessel is connected to the installation for longer time, and it is more susceptible to varying weather conditions). However, the difference in cost may not be significant since a big part of the time required for flushing operations is used for vessel mobilization and connection.

- The new method will require a specialized intervention tool that may have higher renting/maintenance costs and risk of pump failure.

- The new method will require a smaller conduit (smaller injection rates). The load capacity of the vessel might be smaller allowing to use cheaper vessels and enabling faster mobilization and handling.

When flushing using a service vessel and the pump is integrated and part of the subsea system

- Higher initial CAPEX and OPEX than for systems design with the traditional flushing strategy

- The new method will require longer flushing time than the traditional method, which represents higher costs due to vessel rent rates and higher risk (the vessel is connected to the installation for longer time, and it is more susceptible to varying weather conditions). However, the difference in cost may not be significant since a big part of the time required for flushing operations is used for vessel mobilization and connection.

- The new method will require a specialized intervention tool, similar to the one used for standard flushing.

- The new method will require a smaller conduit (smaller injection rates). The load capacity of the vessel might be smaller allowing to use cheaper vessels and enabling faster mobilization and handling.

**Carbon emission accountancy**

Intervention vessels and offshore production systems generate power by using diesel generators and gas turbines, which emit carbon dioxide and other greenhouse gases to air. The new flushing method enables the injection of smaller rates than the traditional single pass method, and thus requires less deck pumping power, which could translate into lower carbon emissions. However, flushing must be run for longer time than in the traditional method, which could translate into higher carbon emissions. To study this issue more in detail, an expression to estimate the total carbon emissions ($CO2$) generated by the deck pump during flushing operations is shown in Eq. 1.

$$CO2_{injection\ pumping} = \frac{\Delta P \cdot Q}{\eta_{deck\ pump}} \cdot t_{flushing} \cdot F_{CO2/energy} \qquad \text{Eq. 1.}$$

In this equation, $\Delta P$ is the pressure boost provided by the pump, Q is the injection rate, $\eta_{deck\ pump}$ is the overall efficiency of the deck pump, $t_{flushing}$ is the total flushing time required and $F_{CO2/energy}$ is a factor that provides amounts of CO2 released by the power generation source per units of energy generated.  Consistent units must be used in this equation.

To determine when the new flushing method to have less carbon emissions than the traditional method ($CO2_{trad\ method} \geq CO2_{new\ method}$) Eq. 1 is substituted both sides. Simplifying terms and assuming that the pressure boost provided by the deck pump is the same for the traditional configuration and the new configuration gives:

$$Q_{trad\ method} \cdot t_{flushing,trad\ method} \geq Q_{new\ method} \cdot t_{flushing,new\ method} \qquad \text{Eq. 2.}$$

The injection flow rates required in the new method are smaller than the one required in the traditional method (based on the experimental data presented earlier, by ½ or ¼). However, the time required to flush is significantly higher. To find out the relationship between required flushing time and injection rate of the new method, Eq. A-6 is used.

$$\alpha_{oil} = \alpha_{oil,initial} \cdot e^{-\tau} \qquad \text{Eq. 3.}$$

Assuming that all the domain is initially filled with oil, then $\alpha_{oil,initial} = 1$. If the desired leftover amount of oil in the geometry is set to 1% or the original volume, and dimensionless time $\tau$ is cleared from the expression, this gives:

$$\tau = -ln(0.01) = 4.6 \qquad \text{Eq. 4.}$$

From the definition of dimensionless time (Appendix A), this gives that $\frac{Q_{new\ method} \cdot t_{flushing,new\ method}}{V} = 4.6$. Substituting in Eq. 2 gives:

$$Q_{trad\ method} \cdot t_{flushing,trad\ method} \geq 4.6 \cdot V$$

When this condition is fulfilled, the new method will generate less CO2 emissions than the traditional method.

Other carbon emissions that are relevant to consider are the one related with the propulsion of the vessel. If the new method allows to use smaller vessels, they will most likely require less power for propulsion and will therefore generate less carbon emissions.

**Limitations of the method proposed**

The new method was developed using experimental data for oil-water systems. Results may be different depending on the physical properties of the fluid (e.g. density and viscosity) and their miscibility. For example, the recirculating flowrate to achieve good mixing for MEG displacing gas could be much higher than for liquid-liquid systems. Experiments should be conducted to study other fluids. When the geometry contains gas in addition to liquid, the recirculation pump must be able to handle gas-liquid mixtures. This technology is often expensive. A better option might be to use an ejector (jet pump).

**Methodology**

The following steps were followed:

1. Describing details of a study case
2. Development of a hydraulic model of the study case to compute required deck pump discharge pressure and power and recirculation pump delta pressure and power.
3. Perform sensitivity studies with the model developed in 2 on MEG injection rate and recirculation rate to determine power of deck pump and discharge pressure of deck pump.
4. Estimating required internal diameter, thickness, and total weight of the pipe used for injection for the cases analysed in 3.
5. Provide observations and discussions.

**Proposed Concept.** The concept proposed in this paper for flushing of subsea equipment uses a small-diameter, high-pressure supply line and a subsea deployed tool containing a pump which recirculates the cleaning fluid through the component or system to be retrieved. A pump was used to induce recirculation, instead of an ejector (Jet pump), as used by Valentini (2019).

The main benefit of this method is the potential to reduce the content of hydrocarbons to acceptable levels with smaller flow rates of flushing fluid supplied from the surface. For instance, the surface vessel may provide 10 m³/h of MEG, but the subsea pump deployed may enable to recirculate this system's fluid and achieve the equivalent flow rate of 25 m³/h.

**Fig. 10** presents a schematic representation of the flushing circulation loop proposed for cleaning subsea production systems. A surface pump, $P_1$, injects the desired rate, $Q_1$, of MEG though a deployable line, such as a flexible hose or coiled tubing (CT). The surface-injected fluid provides the power to the motor, $M_2$, of a positive displacement subsea pump, $P_2$. The pump $P_2$ provides the required pressure for the recirculated fluid, at rate $Q_2$, to merge with the surface fluid. This allows the reservoir fluid in the subsea hardware to be flushed out with a higher flowrate, $Q_3$.

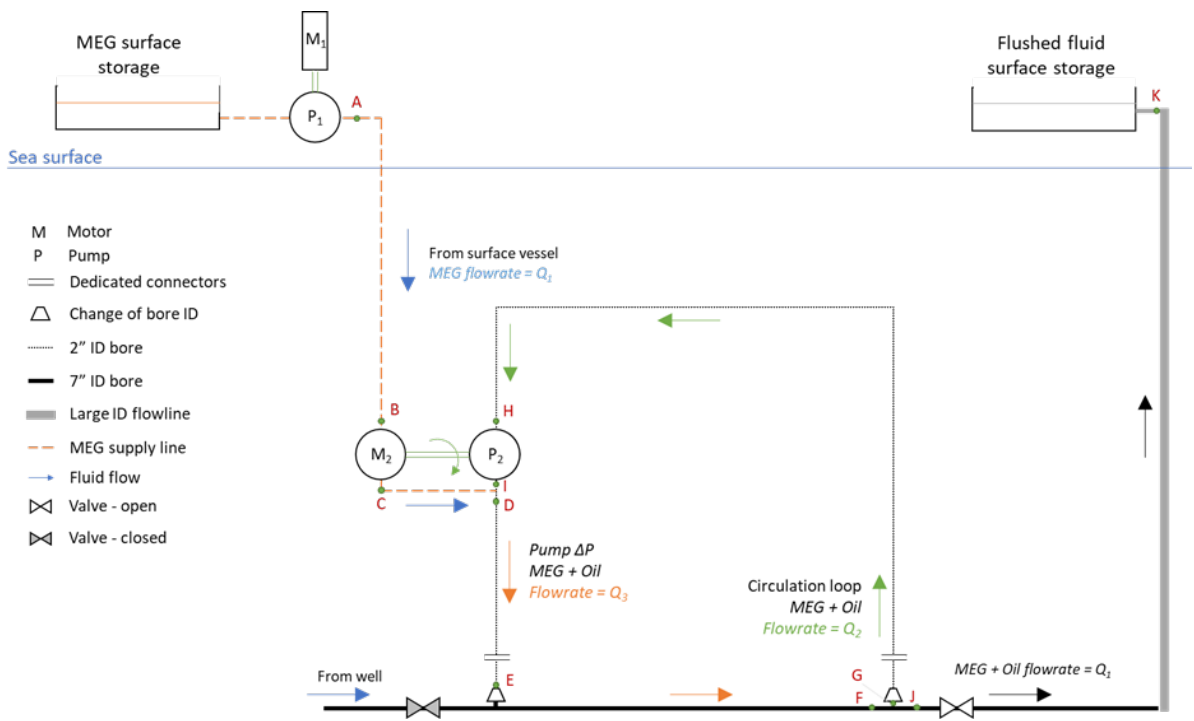

**Fig. 10–Schematics used for pressure loss calculations during circulation of flushing fluid.**

The nodes A through K depicted in Fig. 10 represent the following points in the loop:

A.  Outlet of surface pump, working fluid is MEG supplied from the surface. Flowrate is $Q_1$.

B.  Point on the line of MEG fluid at sea bottom level, at the intake of the motor of the subsea pump.

C.  Point on the line of MEG fluid at sea bottom level, at the outlet of the motor of the subsea pump.

D.  Point at the discharge of the recirculation pump. Mixture of MEG from surface and recirculated fluid. Flowrate is $Q_3$.

E.  Upstream of bore ID transition, from the 2-in recirculation line to the 7-in production line. Flowrate is $Q_3$.

F.  Upstream of T-junction that divides flow in return to surface, at flowrate $Q_1$, and recirculation, at flowrate $Q_2$.

G.  Inlet of branch to recirculation loop, downstream of bore ID transition, from 7-in to 2-in. Flowrate is $Q_2$.

H.  Inlet of subsea pump, mixture of MEG and original fluid in the SPS. Flowrate is $Q_2$.

I.  Outlet of subsea pump, mixture of MEG and original fluid in the SPS.

J.  Fluid returning to the surface. Flowrate is $Q_1$. Pressure losses assumed negligible from this point forward.

K.  Fluid at surface level. Assumed at atmospheric pressure.

Pressure drop calculations were performed along the flow path of the case study to determine technical feasibility of the proposed flushing system and to estimate key operational parameters. The pressure losses in the loop were calculated based on the following expressions, which considers losses due to friction of the flow from one node to another, change of diameter in the pipelines, and pressure variations across pumps and motors. Pressure is denoted by $P$, the subscripts indicate the corresponding system node. Flow velocity is denoted by $V$. Motor and pump efficiency are denoted by $\eta$. Water depths is denoted by $WD$. And $g$ is the gravitation acceleration.

$$P_J = P_K + \rho_{MEG} \cdot g \cdot WD \qquad \text{(Eq. 5)}$$

$$P_F = P_J + \rho_{MEG} \cdot \frac{V_J^2 - V_F^2}{g} \qquad \text{(Eq. 6)}$$

$$P_E = P_F + \Delta P_{E \rightarrow F} + \Delta P_{2"ID \rightarrow 7"ID} \qquad \text{(Eq. 7)}$$

$$P_D = P_E + \Delta P_{D \rightarrow E} \qquad \text{(Eq. 8)}$$

$$P_C = P_D = P_I \qquad \text{(Eq. 9)}$$

$$P_B = P_C + \left(\frac{Q_2}{Q_1}\right) \cdot (P_I - P_H) \cdot \left(\frac{1}{\eta_{M_2} \cdot \eta_{P_2}}\right) \qquad \text{(Eq. 10)}$$

$$P_A = P_B - \rho_{MEG} \cdot g \cdot WD \qquad \text{(Eq. 11)}$$

$$P_H = P_G - \Delta P_{G \rightarrow H} \qquad \text{(Eq. 12)}$$

The pressure at node $G$, $P_G$, must be estimated iteratively. The procedure described by Wang (2011) was used in this study.

**Description of Flushed Subsea Production System.**

For calculations of the pressure losses in the SPS during a flushing operation the following values were used to describe the system:

- Water depth = 2000 m.
- The ID of the MEG injection line was 1.5, 2, 3 and 4-in.
- The ID of the recirculation line was 2-in.
- The ID of the production line was 7-in.
- The length of the recirculation pipeline was 5 m downstream the subsea pump and 10 m downstream the T junction.
- The length of the production line was 10 m upstream the T-junction, and 20 m downstream the T-junction.
- Pressure losses were assumed negligible in the vertical section of the return line to the surface.
- Pumps and motors were assumed to have 100% efficiency.

**Results and Discussion**

**Recirculating Flushing System's Pressure and Power Requirements**.

**Table 2** presents the flowrates assumed during the recirculation flushing operations. Based on Fig. 4, a flow rate of 25 m³/h was deemed sufficient to sweep completely undesired fluids from the subsea component. The rate passing through the subsea component consists of the contributions from (1) the MEG that is injected, flowrate $Q_1$, and (2) the recirculated fluid passing through pump $P_2$, flowrate $Q_2$. Simulations were performed considering different splitting between these two contributions to achieve a total of 25 m³/h. Table 2 shows all cases that were simulated. The injected MEG flow rates are given in the second row of Table 2. The values of recirculated fluid flow passing through pump $P_2$ are given in the third row of Table 2.

Table 2 – Flowrates Assumed in Recirculation Flushing System.

| Case | 1 | 2 | 3 | 4 | 5 | 6 | 7 | 8 | 9 | 10 | 11 | 12 | 13 | 14 |
|---|---|---|---|---|---|---|---|---|---|---|---|---|---|---|
| **MEG Flowrate from Surface, $Q_1$ (m³/h)** | 5 | 6 | 7 | 8 | 9 | 10 | 12 | 14 | 16 | 18 | 20 | 22 | 24 | 25 |
| **Recirculated Flow Rate, $Q_2$ (m³/h)** | 20 | 19 | 18 | 17 | 16 | 15 | 13 | 11 | 9 | 7 | 5 | 3 | 1 | 0 |

The pressure differential supplied by the pump and the required power were calculated according to the flow rate of fluid circulated in the system

**Fig. 11** presents the pressure differential provided by the surface pump, $P_1$, at the given flow rates and selected ID of the MEG injection line. Like Fig. 9, the conventional flushing operation can be compared to the alternative with recirculation. For the injection line with 1.5-in ID, a conventional flushing operation requires the surface pump to provide almost 1000 bar, but this amount could be reduced to 600 bar by having a recirculation flowrate $Q_2$ = 10 m3/h.

If using CT available on the market (Tenaris 2017) for the flushing operation, e.g. the HV-70 line that has an ID of 1.5-in and outside diameter (OD) of 1.75-in and a maximum working pressure of 510 bar, then this would limit the MEG flowrate from the surface to approximatly 12 m³/h, which would not provide a good displacement and cleaning of the subsea production system if using the conventional method. Hence, a CT with a higher ID should be used. For example, a CT of the line HS-70 (Tenaris 2017) with ID 2.0-in and OD 2.375-in has a maximum working pressure of 641 bar. However, using a bigger CT line entails higher costs and complexity.

However, by using the recirculating flushing method it would be possible to achieve the necessary high flow velocities while using the 1.5-in ID injection line (for example, by using the pump $P_2$ to recirculate a $Q_2$ equal to 13 m³/h, while injecting through the CT line $Q_1$ = 12 m³/h). In general, reducing the size of the MEG supply conduit will have a significant impact on the cost, particularly for operations in deep waters.

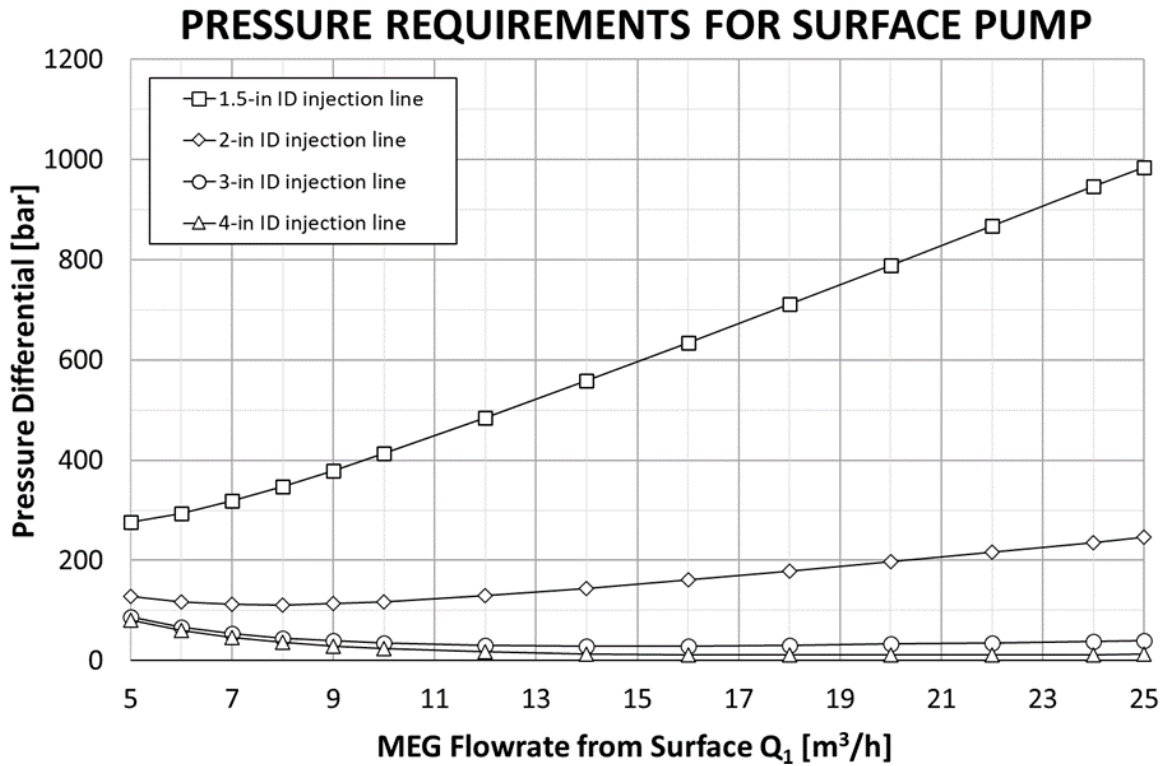

**Fig. 11–Pressure Differentials for Surface Pump, P1.**

As an alternative, a high-pressure flexible steel pipe, such as Coflexip (TechnipFMC. 2006 ), can be used. Coflexip steel pipes can be rated to up to 15000 psi working pressure. Pipes with 2-in or higher ID and rated to 5000 psi, or 345 bar, could be used for single-pass flushing. For a pipe with a 1.5-in ID, it would be required to employ the recirculation system, with flowrates $Q_1 = 8$ $m^3/h$ and $Q_2 = 17$ $m^3/h$.

. **Fig. 12** presents the total power needed to operate the surface pump, $P_1$, at the given flow rates and selected ID of the MEG injection line. Fig. 9 indicates that for the 1.5-in ID injection line, the power needed increases significantly for MEG flowrates greater than 7 $m^3/h$. Fig. 9 may also be used to compare the performance of the flushing system with recirculation, $Q_1 < 25$ $m^3/h$, and the flushing system with a single-pass, $Q_1 = 25$ $m^3/h$. For instance, for the injection line with 1.5-in ID, the single-pass flush requires almost 700 kW, but this amount could be reduced to 250 kW by having a recirculation flowrate $Q_2 = 10$ $m^3/h$.

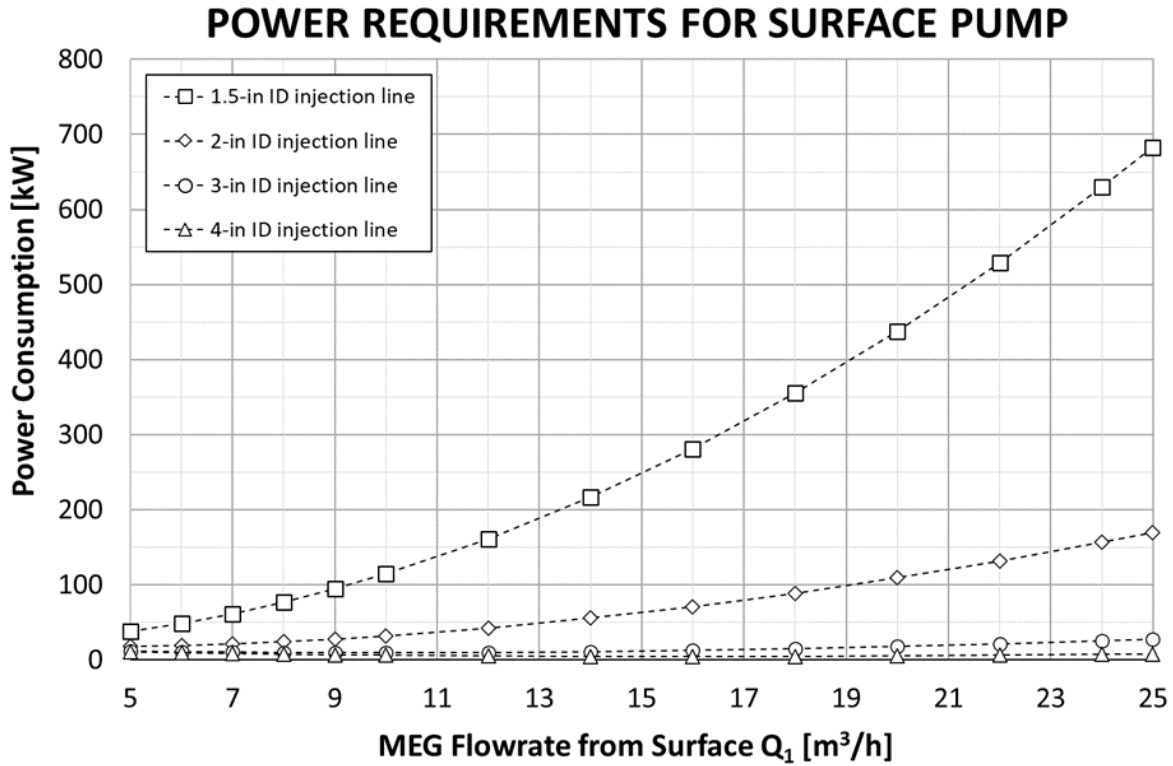

Fig 12–Power Requirements for Surface Pump, $P_1$.

**Reel Drum Storage Capacity**. The dimensions of the reel drum used to store the MEG injection line can be calculated through the following expression:

$$L = (A + C) \cdot (A) \cdot (B) \cdot (K) \qquad \text{(Eq. 13)}$$

where $L$ is the reel capacity, in ft, $A$ is the tubing stack height, in inches, $C$ is the core diameter of the drum, in inches, $B$ is the drum width, in inches, and $K$ is a constant dependent on the OD of the pipe (Drilling Manual 2022). Given that the water depth in the study cases is 2000 m, the desired reel capacity was set to approximately 2400 m, or 7874 ft.

**Table 3** presents results obtained for the reel drum dimensions for the selected CT dimensions. The CT with 1.5-in ID and 1.75-in OD option has a total mass of 7.9 tons, approximately half of the mass of the 2-in ID CT option. This is a significant reduction in storage requirements of the support vessel that would perform the flushing job.

Table 3 – Calculated reel capacity for CT storage.

| ID | OD | K | Core Diameter | Drum Width | Stack height | Reel capacity | Spool volume | Nominal mass | Total mass |
|---|---|---|---|---|---|---|---|---|---|
| [in] | [in] | [-] | [in] | [in] | [in] | [m] | [m³] | [kg/m] | [tons] |
| 1.5 | 1.75 | 0.086 | 50 | 50 | 25 | 2444.6 | 6.4 | 9.6 | 7.9 |
| 2 | 2.375 | 0.046 | 50 | 50 | 39 | 2457.0 | 10.5 | 22.7 | 16.3 |
| 3 | 3.5 | 0.021 | 100 | 50 | 50 | 2444.6 | 25.7 | 47.8 | 31.6 |
| 4 | 4.5 | 0.013 | 100 | 50 | 72 | 2441.9 | 38.3 | 91.0 | 41.3 |

**Table 4** presents results obtained for the reel drum dimensions for the selected Coflexip steel pipe dimensions. Coflexip steel pipes are bulkier than the equivalent CT alternative. The Coflexip steel pipe with 1.5-in ID has a total mass of 53.2 tons, approximately 8 times the mass of the 1.5-in ID CT option. At the same time, the recirculation flushing system enables the use of the Coflexip steel pipe with 1.5-in ID, which would result in at least fewer 23.5 tons of storage needed in the support vessel, when comparing it to the 2-in ID Coflexip steel pipe option.

Table 4 – Calculated reel capacity for Coflexip steel pipe storage.

| ID | OD | K | Core Diameter | Drum Width | Stack height | Reel capacity | Spool volume | Nominal mass | Total mass |
|----|----|----|----|----|----|----|----|----|----|
| [in] | [in] | [-] | [in] | [in] | [in] | [m] | [m³] | [kg/m] | [tons] |
| 1.5 | 3.7 | 0.020 | 38 | 50 | 73 | 2420.3 | 21.8 | 32.4 | 53.2 |
| 2 | 4.5 | 0.013 | 41 | 50 | 93 | 2457.2 | 33.2 | 49.3 | 76.8 |
| 3 | 5.7 | 0.008 | 53 | 100 | 76 | 2409.7 | 54.1 | 80.5 | 119.8 |
| 4 | 7.5 | 0.005 | 69 | 100 | 100 | 2399.3 | 93.1 | 138.6 | 208.2 |

## Conclusions

Experimental and computational results from previous works have shown that a recirculation flushing system using a pump could achieve efficient flushing while using low flow rates of injected flushing fluid. The high flushing efficiency is achieved due to good mixing in the recirculation loop and gradual substitution of the flushed fluid by the injected fluid. The good mixing in the recirculation loop is achieved by using the pump and due to the low pressure drop in the recirculation loop. This method and system have advantages over traditional flushing practices (e.g. single pass) that requires high flow rates to achieve efficient flushing.

This paper has presented an application case of this concept to the flushing of components in subsea production systems, using a recirculating pump subsea deployed by e.g., coiled tubing. The pressure differential and power requirements for the recirculation flushing system were compared to the same requirements for a single pass flushing system.

Calculations for the study cases presented have shown that power requirements and internal pressure could be reduced considerably for flushing systems that employ injection lines with ID of 2-in or smaller. Employing MEG supply lines with smaller an ID would also enable using small capacity vessels for the flushing operation, thus reducing the costs and carbon footprint associated with the operation.

## Nomenclature

$A$ = tubing stack height, in
$B$ = drum width, in
$C$ = core diameter of the drum, in
$g$ = gravitation acceleration, m/s$^2$
$K$ = constant dependent on CT ID, dimensionless
$L$ = reel capacity, ft
$m$ = mass, kg
$P$ = Pressure, Pa or bar
$Q$ = flow rate, m$^3$/s

$V$ = flow velocity, m/s

$WD$ = water depth, m

$\alpha$ = volume fractions, dimensionless

$\Delta P$ = variation of pressure, Pa or bar

$\eta$ = efficiency, dimensionless

$\rho$ = density, kg/m$^3$

## Acknowledgments

This work was carried out as a part of SUBPRO, a Research-based Innovation Centre within Subsea Production and Processing. The authors gratefully acknowledge the financial support from SUBPRO, which is financed by the Research Council of Norway, major industry partners and NTNU.

## Appendix A—Simplified Model to Estimate Variation of Volume Fraction of Oil in Time in the Flushing Geometry

Mass balance in the control volume of **Fig. A-1** yields **Eq. A-1**:

$$\frac{dm_{total}}{dt} = \dot{m}_{in} - \dot{m}_{out} \qquad \text{(A-1)}$$

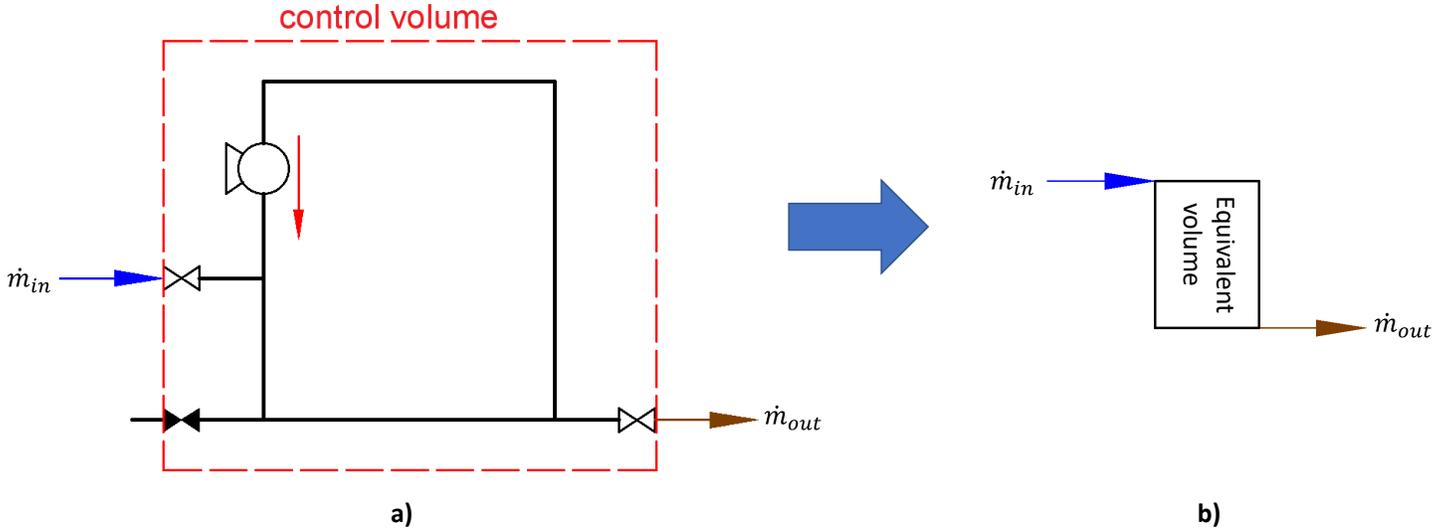

**Fig. A1–Diagram showing a) the configuration of the flushing system using recirculation and a pump, the control volume used to develop the model and the inlet and outlet streams**

Eq. A-1 may be simplified by:

- Expressing the total mass in the control volume as a function of the volume fractions of oil ($\alpha_{oil}$) and water ($\alpha_{water} = 1 - \alpha_{oil}$).

- The only phase that enters the jumper is the flushing liquid (water), at a constant volumetric rate ($q_{water,in}$).

- The stream leaving the jumper has the same volumetric rate as the stream entering the jumper (constant pressure system).

- The recirculation setup ensures that there is good mixing of the phases inside the geometry, thus the stream leaving the geometry will have the same volume fraction values as the geometry.

Thus resulting in **Eq. A-2**:

$$V \cdot \left( \frac{d\alpha_{oil}}{dt} \cdot \rho_{oil} + \frac{d(1-\alpha_{oil})}{dt} \cdot \rho_{water} \right) = q_{water,in} \cdot (\rho_{water} - \alpha_{oil} \cdot \rho_{oil} - (1 - \alpha_{oil}) \cdot \rho_{water}) \qquad \text{(A-2)}$$

The variation of $\alpha_{oil}$ with time may be expressed as:

$$\frac{d\alpha_{oil}}{dt} = -\frac{q_{water,in}}{V} \cdot \alpha_{oil} \qquad \text{(A-3)}$$

Substituting **Eq. A-3** into Eq. A-2, and solving analytically yields **Eq. A-4**:

$$\alpha_{oil} = \alpha_{oil,initial} \cdot e^{-\frac{q_{water,in}}{V} \cdot (t)} \qquad \text{(A-4)}$$

If dimensionless time is used:

$$\tau = \frac{q_{water,in}}{V} \cdot (t) \qquad \text{(A-5)}$$

Then the equation is expressed as:

$$\alpha_{oil} = \alpha_{oil,initial} \cdot e^{-\tau} \qquad \text{(A-6)}$$

Trondheim, October 21$^{st}$, 2024.

**Introduction by Milan Stanko.**

This manuscript was originally submitted to the SPE Production & Operations Journal by Mr. Sevillano in August 29$^{th}$, 2022. The manuscript number assigned was PO-0822-0013.
Reviewer comments and request for major changes were received on November 11th, 2022, with a deadline to submit a new version by May 10$^{th}$, 2023. Unfortunately, it was not possible for Mr. Sevillano to meet this deadline, and his request to change the point of contact to Milan Stanko was unfortunately not answered by the journal.

The response letter to the journal is pasted next, that includes the comments from the reviewer, and technical editors 1 and 2, and answers provided by Milan Stanko (in blue color). To address most of these comments, modifications have been performed on the original manuscript.
The new version of the manuscript will hopefully be published in the same or different journal, or conference sometime in the near future. In the meantime, I welcome people to provide comments either to milan.stanko@ntnu.no, or directly on https://pubpeer.com/

Manuscript PO-0822-0013

Response to Reviewers
Dear Associate Editor of the SPE Production and Operations Journal,

Thank you for giving us the opportunity to submit a revised draft of the manuscript "**A New Method for Flushing Of Subsea System**" for publication in the *SPE Journal*. We appreciate the time and effort that you and the reviewers dedicated to providing feedback on our manuscript and are grateful for the insightful comments on and valuable improvements to our paper. We have incorporated most of the suggestions made by the reviewers. Those changes are highlighted within the manuscript. Please see below, in blue, for a point-by-point response to the reviewers' comments and concerns. We have also corrected some typos and rephased some sentences through the article to improve clarity. All page numbers refer to the revised manuscript file with tracked changes.

**Reviewers' Comments to the Authors:**

**Associate Editor**
Associate Editor:
1.      High speed at low rates is very contradictory. I think this is very misleading. This has to be explained further to be convincing.
**Author response**: Thank you for your comment. The high flow rate in the recirculation loop is achieved by using a pump in the loop.
If the pump is rotor-dynamic (e.g. a centrifugal pump), the recirculation rate will, to a great extent, be independent from the injected flow rate. As long as the pressure drop (pump head) in the recirculation loop is small and the pump is properly sized, it should be feasible to achieve high circulation rates. In a centrifugal pump, the performance map allows for high rates at low heads, or low rates at high heads.
If the pump is a jet pump (ejector), driven by the injected fluid, then it is possible to achieve high recirculation flow rates (e.g. 3-4 times the injected flow rate) as long as the pressure ratio between discharge and suction is small (i.e. small delta pressure), and the ratio of injection pressure to pump suction pressure (rpi in the figure below) is high. A performance map of a jet pump is shown in the figure below.

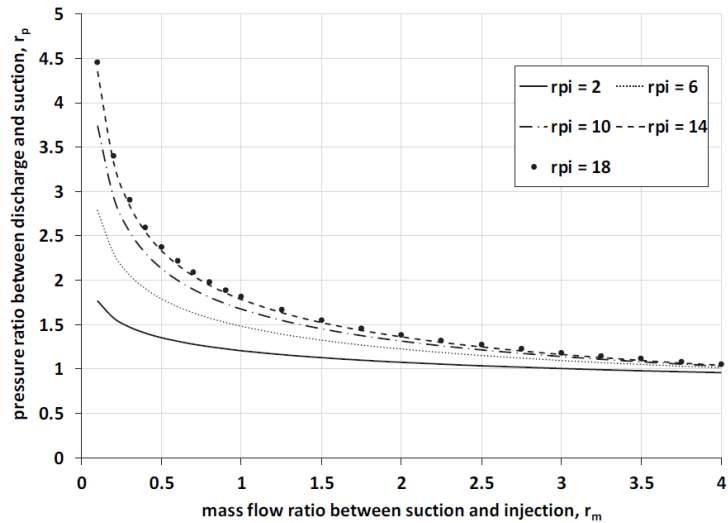

To avoid similar confusion by the readers, we added a short sentence about this in the abstract, in the conclusions.

2. In Figure 7, experimental data basically show that achieving less than 10% is at infinity. It seems like simplified model is somewhat curve fit to data.

**Author response**: Thank you for your comment. By "infinity" you mean at 700 s? The simplified model is not a curve fitting, it is an exponential decline obtained after solving the mass conservation equation in the flushed volume. Coincidentally, it gives a very good match to the experimental data, and because of this, it almost seems like a curve fitting (but it is not).

3. MEG discharge rates to environment is also regulated by local regulations. Paper gives false understanding that MEG discharge to environment is acceptable.

**Author response**: Thank you for your comment.
MEG is included in the OSPAR[1] "List of Substances Used and Discharged Offshore which Are Considered to Pose Little or No Risk to the Environment (PLONOR) – Update 2021"[2], or, according to the Color Categories of the Novatech CHEMS Database, it is considered a green chemical, and, as such, it is allowed to discharge without permission (but reporting is often required). Despite this, O&G companies do not release MEG unless they have to, because of the cost (the cost of MEG per liter is the same as car Diesel). A comment has been added to the introduction to clarify this issue.

4. Configuration in Fig 8 will never reach complete flushing due to recirculation. It will always leave some amount of hydrocarbon in. What is the acceptable level of hydrocarbon left behind?

**Author response**: Thank you for your comment. This is a very valid point. We are afraid we don't have a straight answer.
Any flushing method will most likely leave behind some amount of hydrocarbons in the geometry, most probably in locations like recesses and high points. This doesn't necessarily indicates that the remaining hydrocarbons will be released to sea when the component is retrieved, since it is usually difficult for the remaining small amounts of hydrocarbons to flow out of the geometry.
In Norway, the intentional release of hydrocarbons to sea is usually outside the Emission permit agreement of oil and gas installations. Therefore, in principle, there are no minimum acceptable levels of hydrocarbons that can be released to the sea. If unintentional emissions, this must be reported in the Yearly report of the field to The Norwegian Environment Agency.
In the case of a planned operation such as flushing subsea systems that will release hydrocarbons to Sea, the company must usually apply for a special permit to The Norwegian Environment Agency. The authors were not able to find applications for this kind of permit by doing a web search in relevant governmental websites. However, the authors are aware that there have been some replacements in the last 10 years of subsea equipment in the Norwegian Continental shelf. The fact that we found no applications for release of hydrocarbons could mean that most companies declare there will be no release of hydrocarbons to Sea from flushing operations, despite the fact there will most likely be some hydrocarbon left in the geometry. The new method proposed in our article is based on homogenizing (mixing) hydrocarbons with the flushing fluid (by means of a pump) and then gradually replacing the hydrocarbons by injecting flushing fluids. So, at the end of the flushing time, the geometry will most likely be filled with flushing fluid with traces of hydrocarbon. If the disconnection of the equipment is done right afterwards the flushing is completed, without letting it segregate (MEG is heavier than hydrocarbons), there is a chance there will be some hydrocarbon leaking out of the geometry together with the MEG. The changes of this happening seem to be higher for our method than for the standard single pass flushing method. Workarounds to address this situation could be to wait some time after the flushing is completed, or to circulate for longer time, to reduce the amount of hydrocarbon to a minimum.

---

[1] Convention for the Protection of the Marine Environment of the North-East Atlantic
[2] https://www.ospar.org/documents?d=32939

To make the reader aware of this potential downside of our method, we included a discussion about this in the section "Comparison between the proposed method and the standard flushing method".

5. Pros and cons of both flushing techniques need to be documented and compared.

**Author response**: Thank you for your comment. To address this comment, a new section has been added before the methodology, called: "Comparison between the proposed method and the standard flushing method".

6. Flushing efficiency need to be defined for both techniques.

**Author response**: Thank you for your comment. In this work, the term "flushing efficiency" refers to the volume fraction of hydrocarbons in the geometry after the flushing process. A high flushing efficiency flushing is when that volume fraction is low (close to zero). A poor flushing efficiency is when that volume fraction is high (close to 1).. This sentence has been added to the end of the introduction, before the section "Background – Previous studies"

Technical Editor 1:
The authors present a simple yet clever method for effective flushing of subsea equipments (such as jumpers) not by going to high flowrates, but by smart use of recirculation loops. They demonstrate their approach with lab scale experiments, provide a framework for scaling it up to real world applications and illustrate with examples the significant reduction in opex and load on offshore systems that can be realized. I recommend this article for publication in SPE P&O with some revisions, mostly minor.

**Author response**: Thank you for your comment.

1. The authors clearly and quantitatively establish the benefits of flushing with recirculation loops. For completeness, I recommend including discussion on cost / limitations of flushing with recirculation loops. For example, what would be the additional cost / impact of laying the subsea pumps and recirculation loops (piping & valves)? What would be additional duration of flushing?

**Author response**: To address this comment, two new sections have been added before the methodology, called: "Comparison between the proposed method and the standard flushing method" and "Cost comparison"

2. The CSTR / mixing tank model employed maybe too simplistic for this system. It particularly will be insufficient when applied to systems with larger mean residence times (such as longer pipelines). As a minimum, consider discussing the limitations of the model employed. Axial dispersion models have been known to work well for single pass flushing through pipelines, which may be adapted for systems with recirculation loops. (References: (a) Taylor, G. I. (1954). The dispersion of matter in turbulent flow through a pipe. Proceedings of the Royal Society of London. Series A. Mathematical and Physical Sciences, 223(1155), 446-468. (b) Austin, J. E., & Palfrey, J. R. (1963). Mixing of miscible but dissimilar liquids in serial flow in a pipeline. Proceedings of the Institution of Mechanical Engineers, 178(1), 377-389.)

**Author response**: Thank you for your comment and your suggestion. We agree with the statement that the model is too simplistic for long systems with long residence time. However, we believe the recirculating method/system proposed will have less applicability in large geometries such as long flowlines and pipelines because it will be difficult to achieve high recirculating flow rates with portable pumps. Based on the comparison with laboratory data, the simplified model had a fair accuracy for small geometries, which is the target of the new method. To address this comment, we added some sentences detailing the limitations of the simple model (after fig. 6) and added the references provided by the reviewer. Also, we added an item "applicability for long pipelines" to "Table 1. Qualitative comparison between the traditional flushing method and the new method proposed."

3.      The mixing model currently seems to have no bearing on the proposed concept for subsea flushing. It is not clear to me what is the purpose of this model in relation to the proposed concept. It would be beneficial (for the readers) to see how this model can be useful in practice.

**Author response**: Thank you for your comment and your suggestion. Although simple, the model seems to predict well the experimental data. Therefore, we believe it could be used to plan flushing operations for the new method, e.g. determine the injection flow rate to use and the time to perform flushing for. A comment about this has been added to the article to the second paragraph below Fig. 6.

4.      Page 11, Line 47: Consider estimating carbon emissions reduction. Offshore systems (especially Deep Water) have onsite gas turbines to generate power to run these pumps required for flushing & recirculation.  https://www.energy.gov/sites/prod/files/2016/09/f33/CHP-Gas%20Turbine.pdf although is not specific to offshore gas turbines, shows emissions to be in 0.5 – 0.7 tCO2/MWh. So, for single pass flushing, a power consumption of 700 kW for 2 hours (assumed) would have associated emissions of about 1 tCO2. In comparison, for recirculating flushing, a power consumption of 250 kW for 4 hours (assumed) would have associated emissions of about 0.7 tCO2. That is about a 30% reduction for the example discussed in page 9. This is without considering additional emissions reduction associated with transportation of lighter / compact equipments. Consider including a similar (or better) discussion in your results and/or conclusion.

**Author response**: Thank you for your comment and your suggestion. This is a great addition to the paper. We added a short discussion about this in the section "Carbon emission accountancy", before the methodology section.

5.      Page 9, Line 50/56: Using the same reference above (with $/kW), I expect it to be feasible to perform a high-level cost comparison between flushing without and with recirculation loops. Consider including a similar (or better) discussion in your results and/or conclusion

**Author response**: Thank you for your comment. We added a section titled "cost comparison" before the methodology section. We found that it was challenging to compare both methods with respect to cost, since there are several aspects to consider and many are uncertain. This chapter presents some aspects to consider regarding cost.

6.      Consider discussing any effects of immiscibility and density differences between the two fluids (undesired fluid and flushing fluid) on overall flushing process

**Author response**: Thank you for your comment. A section called "limitations of the method proposed" was added before the "methodology" section to address these issues.

7.      Consider using non-dimensional time $\tau = Qt/V0$ for figures 2, 3, etc. where $Q[m3/h]$ is flowrate for each experiment, $V0$ is total volume of system

**Author response**: Thank you for your comment. We have added this variation of the equation in Appendix A. We tried to express our charts in terms of dimensionless time, but since displaced volume is constant, it was easier to understand by plotting in terms of time and injected flow rate. We used the concept of dimensionless time in the discussion on carbon emissions (located before the methodology)

8.      For consistency with other figures (which show % of replacing fluid), present figs 6 & 7 in terms of water % instead of oil % (which is the undesired fluid)

**Author response**: (pending) Thank you for your comment.

9.      Page 4, Line 25: Please define "high flushing efficiency"

**Author response**: Thank you for your comment. In this work, the term "flushing efficiency" refers to the volume fraction of hydrocarbons in the geometry after the flushing process. A high flushing efficiency flushing is when that volume fraction is low (close to zero). A poor flushing efficiency is when that volume fraction is high (close to 1). This sentence has been added to the end of the introduction, before section "Background – Previous studies"

10.     Page 4, Line 25: Sentence starting with "Moreover, there is a time …" is unclear. Please expand / rewrite. Is this section relating to "dead volume"?

**Author response**: Thank you for your comment. We have rephrased the sentence to improve clarity.

11.     For consistency between text (Page 4, Line 31 and figure 5), include "(Jet Pump)" next to "Ejector"

**Author response**: Thank you for your comment. We have made this change thorough the paper.

12.     Consider including a cross-section of ejector (showing a nozzle, converging & diverging section).

**Author response**: Thank you for your comment. We have added Fig. 6.

13.     Consider including error bars for experimental data

**Author response**: Thank you for your comment. We added a comment about this in the legend of Fig. 5. The students that measured the data shown in Fig. 6 and 7 already graduated so we unfortunately won't be able to add error bars to the figure. But if our memory serves well, we believe they were very small also (as In Fig. 5)

14.     To illustrate the point made on Page 6, Line 25 (on significant difference in flushing without and with recirculation), consider using a plot with just the experimental data for 20.77 m3/h (green dots in figure 3 and green triangles in figure 5)

**Author response**: Thank you for your comment. We believe we achieve the same effect with Fig. 7 (now Fig. 8) and the text provided before the figure.

15.     Consider removing solid red line from figure 7, as it does not appear to be adding any value

**Author response**: Thank you for your comment. We included this curve because we wanted to compare the new method against the traditional method for the same injection rate. This is discussed in the paragraph before Fig. 7 (now fig. 8). We moved the paragraph before the figure to improve clarity.

16.     Page 6, Line 42: I imagine that the proposed concept can be generalized where P2 can be a standalone pump (electrically driven) without the need for M2 and using flow to transfer power from M2 to P2. Consider including a section around such generalizations of your flushing concept

**Author response**: Thank you for your comment. To address this comment we added a section titled "Possible configurations and layouts of the new flushing concept" before the methodology section.

17.     In figure 8, consider showing the section between E and F as a "black-box" to represent a placeholder for any of subsea components (such as production lines, jumpers) that needs flushing as indicated in the introduction

**Author response**: (pending) Thank you for your comment

18.     Consider redoing figure 8 in 3D, showing the recirculation loop laying on the subsea floor

**Author response**: (pending) Thank you for your comment

19.     Equation (1) is missing frictional pressure losses $f\ L/D\ 1/2\ \rho v\ 2$ term, which maybe non-trivial especially at higher velocities

**Author response**: (pending) Thank you for your comment

20.     Equation (2) has $g$ which makes it dimensionally inconsistent. Is it missing ½?

**Author response**: (pending) Thank you for your comment

21.     Consider defining terms like $\Delta PX{\rightarrow}Y$ (and provide mathematical expressions where appropriate)

**Author response**: (pending) Thank you for your comment

22.     Page 8, Lines 31-32: The length description is somewhat unclear. Consider including labels to specify lengths (like $D \rightarrow E$ is $5m$)

**Author response**: (pending) Thank you for your comment

23.     Page 9, Line 2: Please include method for required power calculation (I presume it was product of flowrate and pressure at $A$?)

**Author response**: (pending) Thank you for your comment

24.     Consider swapping the order of Figures 9 and 10 (since Pressure is computed first, and then the power)

**Author response**: Thank you for your comment. We have done this change.

25.     Page 13, Line 53: Is this necessary? Unit conversion factors are trivial and openly available.

Consider removing i

**Author response**: Thank you for your comment. We have removed the conversion factors.

Technical Editor 2:
In general, the paper is good and fit for SPE Journal style. The paper described about novel subsea flushing system which uses a subsea tool to improve the performance of the flushing operation. This flushing operation is typically performed from the host offshore facility through the existing flowlines for chemical inhibition/ MEG, from a service vessel, or a combination of both. The operation can be costly since it will hire vessel and other logistic.

**Author response**: Thank you for your thorough comments, and your time.

Please find here the important observations and major changes from paper is required from the Author:
1). The paper present novelty, good technical information benefit to a practicing engineer  2). However, the structural writing need to be improved and more organized and re-write.  3). The title is advised to be changed to: " A New Method for Flushing of Subsea Production Systems Pre-Decommissioning"

**Author response**: Thank you for your comment. We changed the title of the paper from "A New Method for Flushing Of Subsea System" to "A New Method for Flushing Of Subsea Production Systems Prior to Decommissioning or Component Disconnection"

4). The body of literature is un-organized and need to be re-written, this shall be including the methodology.

**Author response**: Thank you for your comment. We have made changes to the introduction, the background work and to the methodology to address your comment.

5). The step by step of process/ methodology need to be written in structural way and detail. Moreover, conclusion need to be more precise than generic. The author is advised to tabulate "previous research" compare to "his novelty" to keep easy reading for our engineers/ reader

**Author response**: Thank you for your comment. To address this comment, we have made changes to methodology part and we re-wrote parts of the conclusion.

6). One of most important figure is Fig. 8, where the pressure loss calculation is made to support flushing fluid and the relationship to flushed subsea production systems (Wang, 2011) to Table 1 and follow to Figure 9, 10.

**Author response**: (pending) Thank you for your comment. Sorry, We don't understand your  comment.

7). PDF annotations:

**Author response**: Thank you for your thorough comments. We did the following modifications to the manuscript to address your comments:
- Changed the title of the paper from "A New Method for Flushing Of Subsea System" to "A New Method for Flushing Of Subsea Production Systems Prior to Decommissioning or Component Disconnection"

- Moved the last two sentences from the first paragraph in the Abstract to the end of the second paragraph in the abstract.
- Deleted the section title "Methodology" located before "Previous Studies – Background"
- Changed the section subtitle "Previous Studies – Background" to "Background – Previous Studies"
- Inserted the section title "Methodology" before the section "Proposed Concept"
- Inserted a line break after the subtitle "Proposed Concept"
- Changed the title of the results section to "Results and Discussion"
- Inserted a line break after the subtitle "Recirculating Flushing System's Pressure and Power Requirements."
- Re-writing the first part of the conclusion to make it clearer